\documentclass[aps,pra,twocolumn,groupedaddress,showpacs]{revtex4-1}
\usepackage{graphicx, placeins}

\begin{document}
\title{In Situ Momentum Distribution Measurement of a Quantum Degenerate Fermi Gas using Raman Spectroscopy}
\author{Constantine Shkedrov, Gal Ness, Yanay Florshaim, Yoav Sagi}
\email[Electronic address: ]{yoavsagi@technion.ac.il}

\affiliation{Physics Department, Technion - Israel Institute of Technology, Haifa 32000, Israel}

\date{\today} \begin{abstract}
The ability to directly measure the momentum distribution of quantum gases is both unique to these systems and pivotal in extracting many other important observables. Here we use Raman transitions to measure the momentum distribution of a weakly-interacting Fermi gas in a harmonic trap. For narrow atomic dispersions, momentum and energy conservation imply a linear relation between the two-photon detuning and the atomic momentum. We detect the number of atoms transferred by the Raman beams using sensitive fluorescence detection in a magneto-optical trap. We employ this technique to a degenerate weakly-interacting Fermi gas at different temperatures. The measured momentum distributions match theoretical curves over two decades, and the extracted temperatures are in very good agreement with the ones obtained from a conventional time-of-flight technique. The main advantages of our measurement scheme are that it can be spatially selective and applied to a trapped gas, it can be completed in a relatively short time, and due to its high sensitivity, it can be used with very small clouds.
\end{abstract}
\maketitle

In recent years, experiments with ultracold dilute gases contributed immensely to our understanding of quantum many-body phenomena \cite{RevModPhys.80.885}. Among the properties that make these systems so useful are the tunability of inter-particle interactions in the vicinity of a Feshbach resonance \cite{RevModPhys.82.1225}, the flexibility of generating different potentials using far-off-resonance light \cite{Grimm200095}, and the ability to directly measure observables not easily accessible in other systems, such as the momentum of the particles. The momentum distribution is instrumental for extracting many other important observables. For instance, the total energy of the cloud can be calculated from the momentum distribution measured after the interaction energy has been converted into kinetic energy in a ballistic expansion \cite{Ensher1996,DeMarco1999,Bourdel2003,PhysRevLett.97.220406}. In addition, the contact parameter, $C$, can be extracted from the tail of the momentum distribution \cite{PhysRevLett.104.235301}, which scales as $n(k)\rightarrow C/k^{4}$ for $k\gg k_F$, with $k_F$ being the Fermi wave-vector \cite{Tan08}. The occupied single-particle spectral function can also be reconstructed from a set of momentum distribution measurements of rf out-coupled atoms \cite{Stewart2008}. This so-called momentum resolved photoemission spectroscopy (MRPES) technique is one of the most powerful methods to characterize the many-body state of the system \cite{Gaebler2010,Feld2011,Koschorreck2012,PhysRevLett.114.075301}.

The most common way to measure the momentum distribution is by abruptly closing the confining potential and letting the atoms expand ballistically without collisions, commonly referred to as time-of-flight (TOF) measurement. To eliminate systematic errors due to the initial size of the cloud, the expansion time must be very long or alternatively be performed in a perfect harmonic trap for a quarter of the trapping period \cite{PhysRevLett.89.270404,PhysRevLett.100.090402,PhysRevLett.105.230408}. These techniques do not allow determination of the momentum distribution of a gas which is still trapped. Since they rely on optical imaging, they are naturally limited by the relatively small atomic absorption cross-section and the spatial optical resolution.

An alternative approach is to use stimulated Raman scattering. In this process, two different internal states are coupled by the absorption and stimulated emission of two photons from two different optical beams. Conservation of energy and momentum restrict the possible velocity of the atom coupled by the Raman beams. The velocity-selective nature of the Raman process has been employed for manipulation \cite{Kasevich1991a}, velocity detection \cite{Kasevich1991a,Afek2017}, cooling \cite{Aspect1988,Kasevich1992}, and atomic interferometry \cite{Kasevich1991}. Raman coupling was also used for creating spin-orbit coupling for neutral Bose \cite{Lin2011} and Fermi gases \cite{Cheuk2012,PhysRevLett.109.095301}. Bragg scattering, a closely-related process, was used to study the structure factor and excitation spectrum of Bose-Einstein condensates (BEC) \cite{Stenger1999,Steinhauer2002} and degenerate Fermi gases \cite{PhysRevLett.101.250403}, and to measure the momentum distribution of a homogeneous weakly-interacting BEC \cite{Gotlibovych2014}.

In this paper, we demonstrate an \textit{in situ} measurement of the momentum distribution of a degenerate weakly-interacting Fermi gas using Raman spectroscopy and a sensitive fluorescence detection scheme \cite{Shkedrov2018}. There are four main advantages to this approach. First, it can be applied to the atoms without releasing them from the trap. This is particularly useful if a measurement of only part of the cloud is needed, in which case the Raman beams can be focused and probe this part specifically. Second, since our detection scheme is done by recording the fluorescence of atoms trapped in a MOT, it is very sensitive and can be applied with very small atom numbers. Third, the measurement is not sensitive to the shape of the cloud. Lastly, unlike TOF techniques which require an expansion of tens of milliseconds, the measurement using Raman transition can be performed in a short duration, typically much less than a millisecond. This merit can open the door to novel studies of out-of-equilibrium dynamics.

\begin{figure*}
	\makebox[\textwidth][c]{\includegraphics[width=16cm]{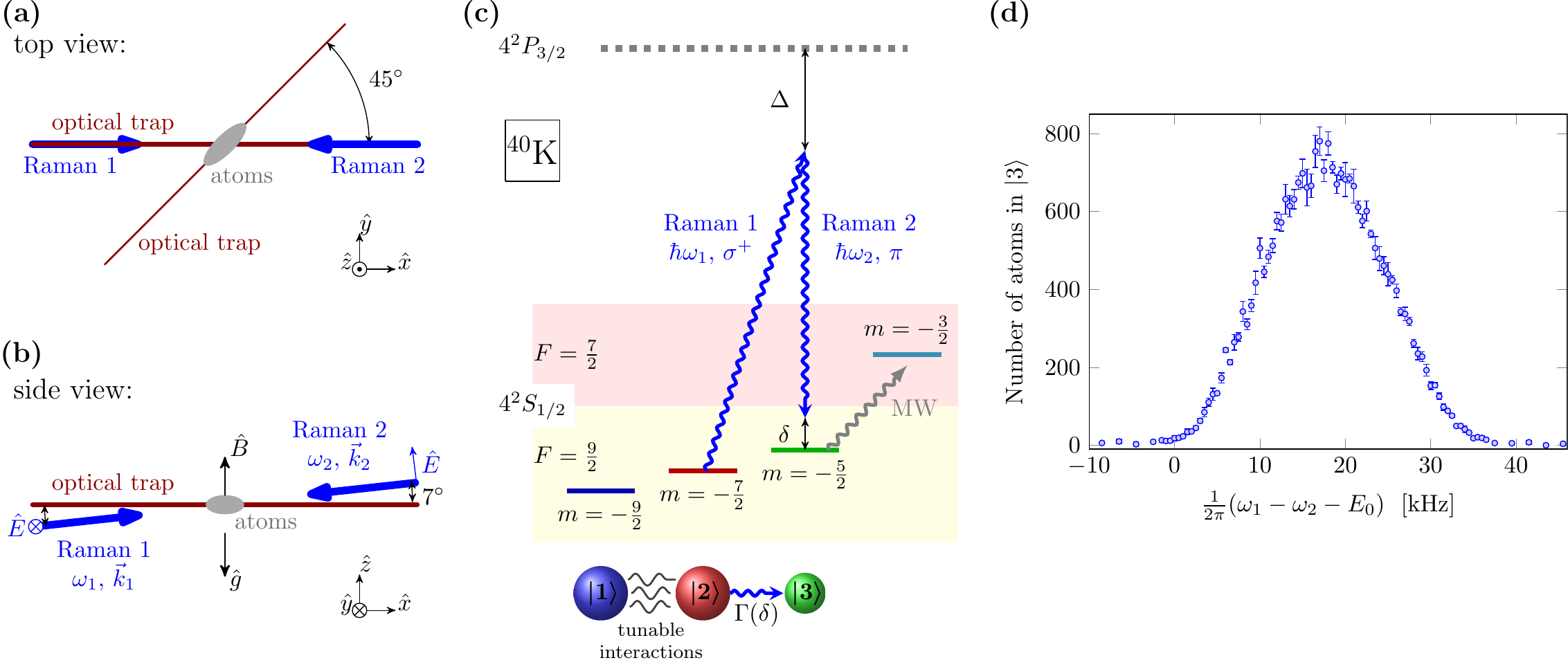}}
	\caption{Top (\textbf{a}) and side (\textbf{b}) schematic plots of the beam configuration in the Raman spectroscopy experiments. Two Raman beams (blue lines with arrows) with orthogonal polarizations are counter-propagating and cross the atomic cloud, which is being held in an elongated optical dipole trap (red lines). The magnetic field axis is parallel to gravitational acceleration. \textbf{(c)}, Level diagram of the relevant states in $^{40}$K. The two states whose interaction is controlled using the Feshbach resonance are designated by $|1\rangle$ and $|2\rangle$. In Raman spectroscopy we measure the momentum-dependent transition rate $\Gamma(\delta)$ from state $|2\rangle$ to the state $|3\rangle$, which is initially unoccupied and is weakly-interacting with atoms in other states, in the range of magnetic fields applied in these experiments. We employ a sensitive fluorescence detection scheme to count the Raman-coupled atoms in state $|3\rangle$ \cite{Shkedrov2018}. First, we use MW adiabatic sweep to transfer these atoms to state $|F=7/2,m_F=-3/2\rangle$, which is magnetically trappable. Then, we trap them by switching on a magnetic quadrupole field. Finally, we switch on 3D MOT beams and record their fluorescence. \textbf{(d)}, Representative raw data recorded in a field of $B=209$G with a $400\mu$s Raman pulse duration and detuning of $\Delta\approx-2\pi\times3.66$GHz. Each data point was repeated 4 times, and the error bars are the $1\sigma$ standard error.}
	\label{fig:sketch1}
\end{figure*}

The two-photon Raman transition connects atoms with energy $E_i(\vec{k})$ and $E_f(\vec{k}+\vec{k}_r)$, in the initial and final states, respectively, where $\hbar\vec{k}$ is the initial momentum of the atom and $\vec{k}_r=\vec{k}_1-\vec{k}_2$ is the relative wave-vector of the Raman transition. Assuming a quadratic dispersion relation $E_j(\vec{k})=E_j^0+\hbar^2 \vec{k}^2/2m$, with $E_j^0$ being the zero momentum energy, conservation of energy and momentum yields:
\begin{equation}
\hbar\delta=k_r\frac{\hbar^2}{m} k_z+\frac{\hbar^2}{2m}\vec{k}_r^2 \ \ ,\label{eq_raman_1}
\end{equation}
	where $k_z\equiv \vec{k}\cdot\hat{k}_r$ is the atomic momentum projection along the direction of $\vec{k}_r$, $\delta=\omega_1-\omega_2-(E_f^0-E_i^0)/\hbar$ is the Raman detuning. Since $\omega_1-\omega_2\ll \omega_1,\omega_2$, we can write $|\bar{k}_r|=\frac{4\pi}{\lambda_R}\sin{\theta/2}$ where $\lambda_R$ is the average wavelength of the Raman beams and $\theta$ is the angle between them. Eq.(\ref{eq_raman_1}) shows there is a linear relation between the Raman detuning and $k_z$. Note that in this treatment we assume a linear response of the Raman transition, which means the Raman pulse duration should be much shorter than the Raman Rabi duration $2\pi\Omega_R^{-1}$. Thanks to the high sensitivity of our fluorescence detection scheme, we can indeed work in the weak coupling regime where the fraction of transferred atoms is typically less than few precents.

\begin{figure}
	\centering
	\includegraphics[width=1\linewidth]{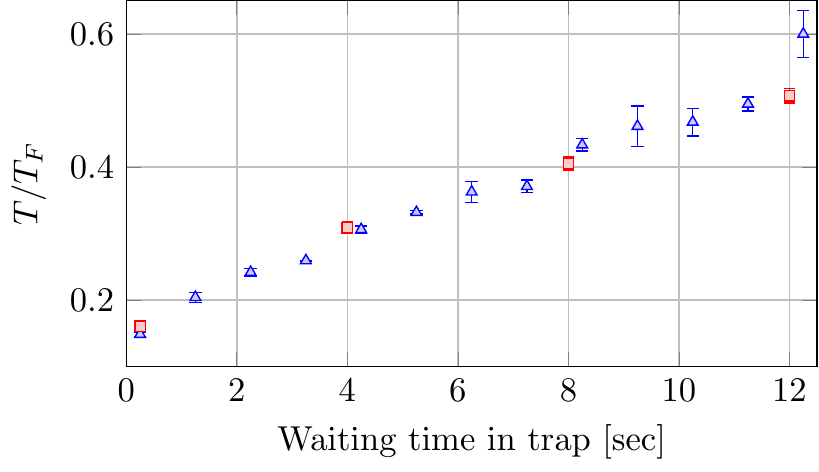}
	\caption{The normalized temperature, $T/T_F$, of a weakly-interacting Fermi gas as a function of waiting time in the optical trap before detection. The red squares are extracted from the one-dimensional momentum distributions measured by Raman spectroscopy, while the blue triangles are extracted from  momentum distributions measured using TOF technique.}
	\label{fig:waiting_time_in_trap}
\end{figure}

The experiments are conducted with a quantum degenerate gas of fermionic $^{40}$K atoms. The gas is prepared in a balanced incoherent mixture of the states $|1\rangle=|F=9/2,m_F=-9/2\rangle$ and $|2\rangle=|F=9/2,m_F=-7/2\rangle$ in the $4^2\textrm{S}_{1/2}$ manifold, whose interaction, characterized by the s-wave scattering length $a$, can be tuned by an external magnetic field near a Feshbach resonance at $B_0=202.14$G \cite{Shkedrov2018}. The experimental apparatus and cooling sequence are the same as described in Refs. \cite{Shkedrov2018,Ness2018}. Two counter-propagating Raman beams overlap with the atomic cloud (see figure \ref{fig:waiting_time_in_trap}a and \ref{fig:waiting_time_in_trap}b). The Raman process, shown schematically in figure \ref{fig:sketch1}c, couples atoms in the state $|2\rangle$ to a third initially unoccupied state $|3\rangle=|F=9/2,m_F=-5/2\rangle$. We denote the Raman beams frequencies by $\omega_1$ and $\omega_2$ and their wave-vectors by $\vec{k}_1$ and $\vec{k}_2$. The detuning of $\omega_1$ and $\omega_2$ from the closest transition, $4^2\textrm{S}_{1/2}\rightarrow4^2$P$_{3/2}$, is given by $\Delta$, which is chosen to be much larger than the excited state width in order to keep a low rate of spontaneous Raman scattering events.

The Raman beams are derived from a single distributed Bragg reflector (DBR) laser whose wavelength can be tuned by changing the temperature and current. The linewidth of the laser is $\sim 1$MHz, and with temperature stabilization it is stable to within $10$MHz, much smaller than $\Delta$. The two beams are generated by two acousto-optical modulators in a double-pass configuration \cite{Donley2005}. The modulators are driven by a single direct digital synthesizer (DDS), which ensures that the two Raman beams are phase-coherent. The beams are delivered to the apparatus by two single-mode polarization-maintaining optical fibers. Their $1/e^{2}$ radius is $0.9$mm and their power is around $1$mW. They have linear orthogonal polarizations, and their propagation axis forms an angle of $\sim 83^\circ$ with the direction of the magnetic field (the quantization axis) and $45^\circ$ with the long axis of the optical trap (see figure \ref{fig:sketch1}a and \ref{fig:sketch1}b). With this choice, we are able to drive the Raman transition $|2\rangle\rightarrow |3\rangle$ which requires $\sigma^+$ and $\pi$ photons.

In order to achieve the highest detection sensitivity, we employ a technique we have recently developed for rf spectroscopy \cite{Shkedrov2018}. The main idea is to selectively transfer only the atoms in state $|3\rangle$ to another magnetically trappable state which is initially unoccupied. Then, these atoms are trapped by turning on a magnetic trap, while driving atoms in all the other untrappable states out of the detection region. We then turn on laser beams which create a 3D MOT in which we detect the trapped atoms by recording their fluorescence. This technique allows us to detect very small signals down to only few atoms.

The Raman spectrum is obtained by scanning the two-photon detuning, $\omega_1-\omega_2$, and recording the number of atoms transferred to state $|3\rangle$. A characteristic measurement is shown in figure \ref{fig:sketch1}d. Each data point requires that the cloud be prepared anew. For each spectrum, we also take several data points at $\delta=-2\pi\times 500$khz, far from the Raman resonance around $\sim2\pi\times 18$kHz (see figure \ref{fig:sketch1}d). These shots are used as a calibration of the background signal, which originates from spontaneous Raman scattering events. The background, which is typically 20-40 times smaller than the peak Raman signal, is subtracted from all measurements presented in this paper.

\begin{figure}
	\centering
	\vspace{5mm}
	\includegraphics[width=1\linewidth]{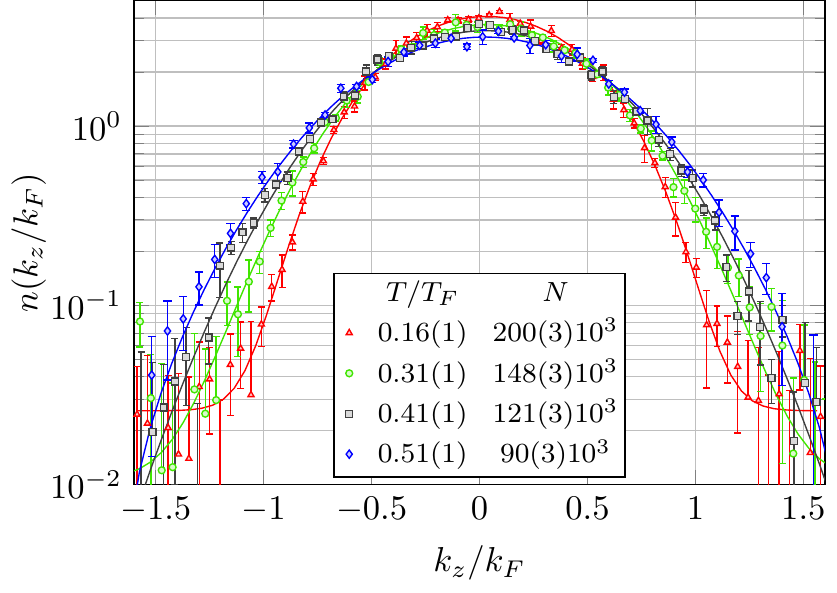}
	\caption{The one-dimensional momentum distribution of a spin-balanced weakly-interacting Fermi gas measured by Raman spectroscopy. Each graph corresponds to a different waiting time in the optical trap before detection, and therefore also to different $T/T_F$. The waiting times are $0.25$ sec (red triangles), $4$ sec (green circles), $8$ sec (black squares) and $12$ sec (blue diamonds). The Raman signal is normalized such that $\int_{-\infty}^{\infty}n(k_z/k_F)d(k_z/k_F)=4\pi/3$. Each point in these graphs is an average over three repetitions and error bars are $1\sigma$ standard errors. The solid lines are fits to the theory of Eq.(\ref{mom_dist_har_1axis}), from which we extract $T/T_F$ shown together with the number of atoms in the inset table. In these experiments, the Raman pulse duration was $500\mu$s, and it was applied $500\mu$s after turning off the optical trap. The single-photon detuning in these measurements is $\Delta=-2\pi\times 54.78(8)$GHz relative to the $|F=9/2,m_F=-9/2\rangle \rightarrow |F'=11/2,m_F=-11/2\rangle$ transition.}
	\label{fig:ramandatainharmonictrap1}
\end{figure}

Raman spectroscopy measurements of the one-dimensional momentum distributions of a weakly-interacting degenerate Fermi gas are shown in figure \ref{fig:ramandatainharmonictrap1}. In these measurements, the atomic cloud was prepared in a magnetic field of $B\approx185$G where $1/k_F a\approx5.7$ and states $|1\rangle$ and $|2\rangle$ are weakly-interacting. After evaporative cooling in the optical trap, we ramp up the trap such that the radial and axial oscillation frequencies are $\omega_r=2\pi\times 750(3)$Hz and $\omega_z=2\pi\times 36(2)$Hz, respectively. In order to prepare a gas with different $T/T_F$ ($T_F$ is the Fermi temperature), we vary the waiting time in the trap before applying the Raman pulse. The two-photon Raman detuning, $\delta$, is translated into momentum using Eq.(\ref{eq_raman_1}) and normalized by $k_F$, which is calculated from the measured number of atoms and the trap oscillation frequencies. The quantity $(E_f^0-E_i^0)/\hbar$, namely the energy difference between the initial and final states, is measured by rf spectroscopy with a spin-polarized gas \cite{Shkedrov2018}.

One-dimensional momentum distribution of non-interacting fermions in a harmonic trap has the following form:
\begin{equation}
n(k_z)=-8\sqrt{\pi}\left(\frac{T}{T_F}\right)^{5/2}\mathrm{Li}_{5/2}\left(-\zeta e^{-\frac{k_z^2/k_F^2}{T/T_F}}\right) \ \ ,\label{mom_dist_har_1axis}
\end{equation}
where $\mathrm{Li}_n(z)$ is the polylogarithmic function and $\zeta$ is the fugacity. This relation can be obtained by doubly integrating the three-dimensional momentum distribution $n(k)=-8/\sqrt{\pi}\left(\frac{T}{T_F}\right)^{3/2}\mathrm{Li}_{3/2}\left(-\zeta e^{-\frac{k^2/k_F^2}{T/T_F}}\right)$ over the $x$ and $y$ directions \cite{Ketterle2008}. In a harmonic trap the fugacity is related to the normalized temperature through:
\begin{equation}\label{eq_fugHar}
-\mathrm{Li}_3\left(-\zeta\right)=\frac{1}{6}\left(\frac{T}{T_F}\right)^{-3}  \ \ .
\end{equation}

We fit the data with Eq.(\ref{mom_dist_har_1axis}) (solid lines in figure \ref{fig:ramandatainharmonictrap1}). When fitting, Eq.(\ref{eq_fugHar}) constrain the fugacity $\zeta(T/T_F)$, such that there are only $3$ free parameters: $T/T_F$, the distribution center, and a background level, which is typically very small because of our procedure to subtract the background. The excellent agreement between the fits and the data over more than two decades is a compelling evidence that indeed the measurement yields the momentum distribution. In figure \ref{fig:waiting_time_in_trap}, we plot the normalized temperature, $T/T_F$, extracted from the fits (red squares) as a function of the waiting time in the optical trap. For comparison, we also plot the temperature extracted from the momentum distribution measured with a conventional TOF technique (blue triangles). The temperatures extracted by the two techniques agree to within the experimental uncertainty.

In conclusion, we have explored the use of Raman transitions for probing the momentum distribution of a weakly-interacting quantum degenerate Fermi gas in a harmonic trap. There are several ways to extend Raman spectroscopy to the strong interaction regime. If the quantum state supports quasi-particles with a well-defined dispersion relation, e.g. a Fermi liquid, then the considerations leading to Eq.(\ref{eq_raman_1}) may still hold, albeit with a renormalized mass and a mean-field energy shift. The Raman spectrum is then expected to reflect the momentum distribution of the quasi-particles. In the general case, Raman spectroscopy can be used to probe the single-particle spectral function, similar to rf MRPES \cite{PhysRevLett.98.240402,PhysRevA.80.023627}. This, however, requires resolving the momentum distribution of the Raman-coupled atoms. In order to measure the momentum distribution of strongly-interacting fermions, it is necessary to rapidly ramp the magnetic field to the zero-crossing point where $a\approx 0$ ($\sim209$G for $^{40}$K \cite{Smale2019}) \cite{Bourdel2003,PhysRevLett.95.250404}. This ramp has to be faster than the many-body timescale $h/E_F$ \cite{PhysRevLett.104.235301}. Finally, when the Raman spectrum reflects the momentum distribution, it should be symmetric at equilibrium. Indeed, we have observed that in cases where the cloud has not yet reached equilibrium, the Raman spectrum shows interesting asymmetric patterns which are not noticeable in TOF measurements. This sensitivity suggests that Raman spectroscopy can be instrumental in exploring the non-equilibrium dynamics of driven Fermi gases.

This research was supported by the Israel Science Foundation (ISF) grant No. 888418, and by the United States-Israel Binational Science Foundation (BSF), Jerusalem, Israel, grant No. 2014386.


%

\end{document}